\documentclass[amsmath,amssymb,twocolumn,superscriptaddress,10pt]{revtex4-1}

\usepackage{mathtools}

\usepackage{natbib}

\newcommand\G{\mathbf G}
\newcommand\rs{\mathcal{R}}
\newcommand\SE{\boldsymbol\Sigma}
\newcommand\HH{\mathbf H}
\newcommand\SO{\mathbf S}

\newcommand\rsG{\mathbf G^{\rs}}
\newcommand\rsSE{\boldsymbol\Sigma^{\rs}}
\newcommand\rsH{\mathbf H^{\rs}}
\newcommand\rsS{\mathbf S^{\rs}}

\newcommand\kk{\mathbf k}
\newcommand\EE{E}
\newcommand\zz{z}

\newcommand\sisl{\textsc{sisl}}
\newcommand\siesta{\textsc{Siesta}}
\newcommand\tsiesta{\textsc{TranSiesta}}
\newcommand\tbtrans{\textsc{TBtrans}}

\newcommand\Eref[1]{Eq.~\eqref{#1}}
\newcommand\Fref[1]{Fig.~\ref{#1}}

\begin{document}

\title{Removing all periodic boundary conditions: Efficient non-equilibrium Green function calculations}
%\title{Efficient non-equilibrium Green function calculations without periodic boundary conditions}
%			Efficient algorithm for real-space self-energies to remove periodic images in
%    defect and transport calculations}

%\title{Efficient algorithm for real-space self-energies: remove periodic images in
%    defect and transport calculations}

\author{Nick Papior}
\affiliation{Computing Center, Technical University of Denmark, DK-2800 Kongens Lyngby, Denmark}
\affiliation{Center for Nanostructured Graphene, Department of Physics, Technical University of Denmark, DK-2800 Kongens Lyngby, Denmark}
\email{nicpa@dtu.dk}

\author{Gaetano Calogero}
\affiliation{Dipartimento di Ingegneria dell'Informazione, Universit{\`a} di Pisa, 56122, Pisa, Italy}
\affiliation{Center for Nanostructured Graphene, Department of Physics, Technical University of Denmark, DK-2800 Kongens Lyngby, Denmark}

\author{Susanne Leitherer}
\affiliation{Center for Nanostructured Graphene, Department of Physics, Technical University of Denmark, DK-2800 Kongens Lyngby, Denmark}

\author{Mads Brandbyge}
\affiliation{Center for Nanostructured Graphene, Department of Physics, Technical University of Denmark, DK-2800 Kongens Lyngby, Denmark}

\date{\today}

\begin{abstract}
  We describe a method and its implementation for calculating electronic structure and electron transport without approximating the structure using periodic super-cells.
  This effectively removes spurious periodic images and
  interference effects. Our method is based on already established methods readily
  available in the non-equilibrium Green function formalism and allows for non-equilibrium transport. We present examples of a N defect in graphene, finite voltage bias transport in a point-contact to graphene, and a graphene-nanoribbon junction. This method is less costly, in terms of CPU-hours, than the super-cell approximation.
\end{abstract}

\maketitle

\section{Introduction}

Widely used, efficient computational methods have been developed for calculations of electronic structure of systems presenting perfect periodic repetition of a unit cell along one, two, or three dimensions surrounded by vacuum. These are typically based on Kohn-Sham Density Functional Theory (DFT)\cite{Martin2004,Kohanoff2006}. The infinite system is replaced by finite unit-cell with periodic boundary conditions (PBC) using Bloch's theorem and a discrete sampling of Bloch phases/$\kk$-points. Due to the efficient implementations this method is also applied systems which lack periodicity. For example surfaces are modelled by a slab, isolated defects by periodically repeated defects surrounded by ``large'' regions of bulk, and isolated adsorbates on surfaces by a mix. This results in compromises due to computational feasibility with respect to slab-size and inter-defect distances, which may lead to unwanted effects related to interferences or standing-wave patterns not present in the ideal, large system.
Beyond PBC methods have been around for a long time. These include matching of the wavefunctions in different regions, e.g. surface and bulk\cite{APPELBAUM1972}, and Green function or embedding methods\cite{Williams1982,MacLaren1989,Feibelman1992,Ishida1997,Inglesfield2001,Li2018} have e.g. been used to treat the isolated defect/adsorbate on a surface or electronic transport between two electrodes\cite{Wortmann2002}. These methods are based on a screening assumption where the potential has converged to its bulk value outside the computational ``active'' region.

In particular, for transport calculations the treatment of systems as ``open'' with semi-infinite electrodes along the transport directions is essential. A number of computational implementations have been developed for more than a decade for this problem based on the non-equilibrium Green function (NEGF) method\cite{TaGuWa.2001, Brandbyge2002, Thygesen2005, Papior2017,Ozaki2010,Smidstrup2019}. These typically represent electrodes by a unit-cell repeated as periodic layers along the semi-infinite electrode/transport direction, and use PBC and corresponding $\kk$-points in the directions transverse to this. A self-energy is then used to treat the semi-infinite direction in a numerically exact way based on a very efficient method\cite{Sancho1984} which recursively removes the infinite number of degrees of freedom/states in the semi-infinite direction. This approach may also be used in ``single-electrode mode'' treating the surface of semi-infinite bulk
with a computational load comparable to slab calculation of e.g. chemical reactions at the surface \cite{Sanchez-Portal2007,Papior2017,Smidstrup2017}. Indeed this avoids the periodic images and finite size effects of the slabs in the surface-normal direction, but leaves the periodicity in the surface direction.

In this paper we present a simple, efficient and precise method based on Green function theory 
which can be used for isolated defects as well as extended NEGF calculations using multiple probes/electrodes. Our method solves this problem by calculating the real-space self-energy which can be outlined as (details explained in the Method section)
\begin{equation}
    \label{eq:real-space-term}
    \rsSE_{00} = \rsS z - \rsH - [\rsG_{00}]^{-1}.
\end{equation}
Equation~\eqref{eq:real-space-term} is computationally demanding since real-space quantities requires a dense integration grid in reciprocal space.
We emphasize that our method focuses on the efficient algorithmic implementation which has prohibited the community to extend its broader use. Secondly, our method allows non-equilibrium calculations by the regular assumption of ``equilibrium'' electrodes\cite{TaGuWa.2001, Brandbyge2002, Thygesen2005, Papior2017,Ozaki2010,Smidstrup2019}.
The paper is organized as follows. First we describe the theoretical and computational details of our method. Then we show DFT$+$NEGF results using the real-space self-energies for three illustrative cases: i) electronic structure of a nitrogen defect in a large graphene lattice, ii) non-equilibrium transport in a gold STM tip in contact with a graphene flake, and iii) a graphene/graphene nanoribbon junction.

\section{Method}%
\label{sec:method}

We remark that Eq.~\eqref{eq:real-space-term} is a well known equation in the transport community and that our contribution here lies in the implementation.
In the following
we will describe the method for a pristine bulk system which is the basis for defected systems.

\begin{figure}
  \centering
  \includegraphics{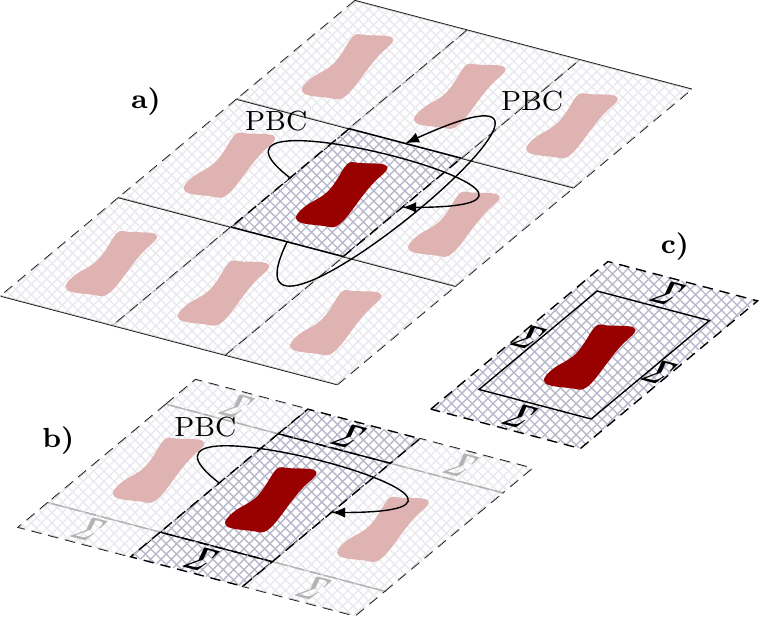}
  \caption{Defected structures using various PBC simulations, neighbour images shown with transparency. a) shows a typical DFT calculation with a single defect (marked region) and PBC in all directions. Using current state of the art NEGF one can remove images in two directions reducing the PBC to 1 direction. Finally in c) our method allows removing all PBC using an enclosing self-energy.}
  \label{fig:problem}
\end{figure}

The starting point of the efficient real-space self-energy method is any system with PBC in 2 or 3 directions where one wishes to replace a predefined direction with a semi infinite description, see Fig.~\ref{fig:problem}a. This may be efficiently described using two semi-infinite directions and one PBC direction, see Fig.~\ref{fig:problem}b, lastly our presented method replaces any number of PBC, and/or semi-infinite directions, with a single self-energy, see Fig.~\ref{fig:problem}c.

The Hamiltonian of an infinitely large system may be written in a block-tri-diagonal ``shell'' fashion:
\begin{equation}
  \label{eq:H}
  \rsH =
  \begin{bmatrix}
    \HH_{00} & \HH_{01} & \mathbf 0 & \dots
    \\
    \HH_{10} & \HH_{11} & \HH_{12} &
    \\
    \mathbf 0 & \HH_{21} & \HH_{22} &
    \\
    \vdots & & & \ddots
  \end{bmatrix}.
\end{equation}
Here index $0$ in $\HH_{00}$ is referred as the \emph{primary} unit cell with only nearest neighbour couplings, $\HH_{01}$ is the coupling between the
primary unit-cell and the first set of neighbour cells (2D: $8$, 3D: $26$), and $\HH_{i,i+1}$ is the coupling between the $i$ and $i+1$ shell. We use the superscript $\mathcal{R}$ to indicate the real-space representation of matrices.

We want to calculate the
Green function for the infinite matrix comprising the Hamiltonian $\rsH$ in a subspace
$\rsH_{0\dots i,0\dots i}\equiv\{\HH_{00},\dots,\HH_{ii}\}$ up to some shell size $i$. The straight
forward Dyson equation is sufficient for systems with short screening lengths such as
metals where the convergence requires only a few shells\cite{WU1995}.
For weak screening the increasing matrix sizes with $i$ in the Dyson equation become problematic and one may replace the real-space iterations in shells with an integral over $\kk$-points to calculate the real-space Green
function (here only shown for the primary unit cell),
\begin{align}
  \label{eq:rs-G}
  \rsG_{00}(\zz) &= \int\!\!\mathrm{d}\kk\, \G_\kk(\zz)
  \\
  \label{eq:rs-SE}
  &= [\rsS \zz - \rsH]_{00} ^{-1} = [\SO_{00} \zz - \HH_{00} - \rsSE_{00}]^{-1}
\end{align}
where $\G_\kk(\zz)$ is the Green function for a given $\kk$-point, $\rsS$ the overlap matrix, and energy plus
imaginary part is $\zz=\EE+i\eta$. We define the subspace of interest by $0$ and the coupling of this to the surrounding bulk system is described by the real-space self-energy, $\rsSE_{00}$. We remark that
$\G_\kk^{\mathrm{T}}=\G_{-\kk}$ using time-reversal symmetry. This converts the inversion
of infinite matrices in real-space to a problem of inverting finite-sized matrices by introducing a $\kk$ integral employing Bloch's theorem. This method was employed in Ref.~\onlinecite{Li2018}.

\begin{figure}
  \centering
  \includegraphics{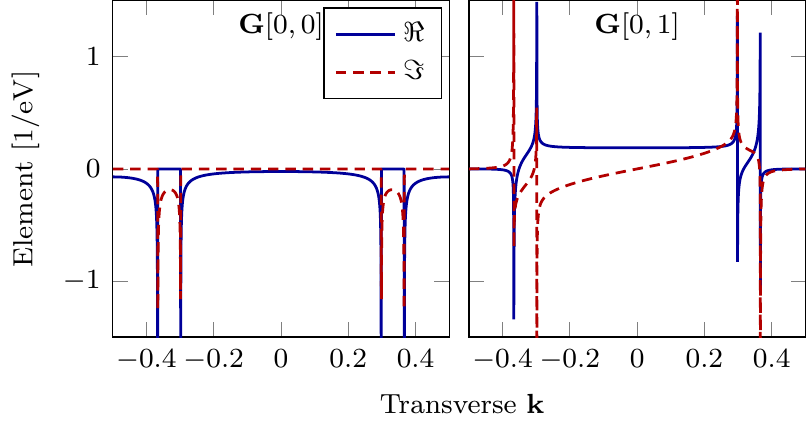}
  \caption{Green function matrix elements of the graphene tight-binding model
      ($t=-2.7\,\mathrm{eV}$) for $\zz=(0.5+i10^{-4})\,\mathrm{eV}$
      ($\G_\kk^{-1}=\mathbf I\zz - \HH_\kk - \SE^L_\kk-\SE^R_\kk$) for transverse $\kk$
      points. Several $\delta$ peaks are seen which makes an integral extremely
      difficult.}
  \label{fig:G-2k}
\end{figure}

Two new problems arise. A sufficient accuracy in the integral is difficult
because the elements of the Green function has Lorentzian peaks/step-functions (in
$\kk$-space) for each eigenvalue (pole). In \Fref{fig:G-2k} we show the Green function
matrix elements (left: diagonal, right: off-diagonal) for fixed
$\zz=(0.5+i10^{-4})\,\mathrm{eV}$ using the standard orthogonal tight-binding model for
graphene with hopping $t=-2.7\,\mathrm{eV}$. We employ the recursion along one direction in graphene to obtain the Greens function as a function of $\kk$ for the direction transverse to this. In effect this means that we sample an extremely dense $\kk$-grid along one direction and a sparser $\kk$-grid transverse to this. We see that the matrix elements comprise both step-functions and convolutions of Lorentzian and step functions. Such functions require dense integration grids to resolve. Note that using Fourier transforms results in the same deficiencies to resolve the peaks.

A second problem is the matrix dimensions of $\G_\kk$. Our interest is to calculate the
real-space Green function in some multiples of the primary unit cell such that the final
matrix has dimensions $n\prod n_{a_i}$ with $n$ being the number of orbitals in the
primary unit cell, and $n_{a_i}$ is the number of repetitions along the $i$'th lattice
vector. For large $n_{a_i}$ the matrix dimensions rapidly increases making a fine integral
in \Eref{eq:rs-G} unfeasible\cite{Settnes2014}.

Our method solves this dimension problem by only performing the costly inversions on
matrices of dimension $n$, regardless of all $n_{a_i}$. This is achieved using three well
established methods and performing them in the following order, 1) a surface self-energy
removes the $\kk$ integral along a single semi-infinite direction, 2) recursive Green
function calculation (block-tri-diagonal inversion, BTD) expands along the semi-infinite
direction and 3) Bloch's theorem efficiently expands the Green function into the remaining
one (or two) dimensions. Note that the surface self-energy calculation is a particular
efficient solution of the BTD algorithm for a bulk system. Hence the difference between
the two methods is subtle, yet significant in their application for our method. The three
steps above work for both surfaces and bulk systems, with minor variations. Note that for
3D systems, leaving a $\kk$ direction out of the integral, \Eref{eq:rs-G}, one finds the
real-space Green function for a cylinder with the directions normal to the cylinder
surface integrated out, retaining the $\kk$ point along the cylinder. Such a particular
use-case will not be covered in this study, but we remark that our code allows such
calculations which may be useful for e.g. line-defects in solids.

In the following all matrices not denoted by $\mathcal R$ are implicitly $\kk$ dependent.

\subsection{Surface self-energies}%

The recursive surface self-energy method by \citeauthor{Sancho85} calculates the left/right self-energies for a
given transverse $\kk$-point. This procedure presents a $2^i$ convergence
series such that for iteration $i$ one have effectively decimated $2^i$
layers. The algorithm is given here for the sake of completeness:
\begingroup
\def\bR{\mathbf R} \def\bL{\mathbf L} \def\btau{\boldsymbol\tau}
\begin{subequations}
  \begin{align}
    \SE^L_0 &= \SE^R_0 = \mathbf 0,
    \\
    \bL_0 & = \HH_{10} - \SO_{10}\zz
    \\
    \bR_0 & = \HH_{01} - \SO_{01}\zz
    \\
    \intertext{perform following iterative scheme until
        $\SE^{L/R}_{i-1}\approx\SE^{L/R}_{i}$:} \btau^L & = [\SO_{00}\zz - \HH_{00} -
    \SE^L_{i-1} - \SE^R_{i-1}]^{-1}\bL_{i-1}
    \\
    \btau^R & = [\SO_{00}\zz - \HH_{00} - \SE^L_{i-1} - \SE^R_{i-1}]^{-1}\bR_{i-1}
    \\
    \SE^L_i & = \SE^L_i + \bL_{i-1}\btau^R
    \\
    \SE^R_i & = \SE^R_i + \bR_{i-1}\btau^L
    \\
    \bL_i & = \bL_{i-1}\btau^L
    \\
    \bR_i & = \bR_{i-1}\btau^R.
  \end{align}
\end{subequations}
\endgroup
The surface self-energy removes the $\kk$ integral along the semi-infinite direction and
immediately reduces the integral in \Eref{eq:rs-G} by one dimension.

\subsection{Block-tri-diagonal inversion (BTD)}%
\label{sec:btd}

This method may be generalized to calculate the layer off-diagonals for the inverse of matrices when these can be written in block form \Eref{eq:H} \cite{Papior2017}. The pristine bulk system may be written in the
following BTD form along the semi-infinite direction:
\begin{equation}
  \label{eq:BTD}
  \HH =
  \begin{bmatrix}
    \HH_{00} & \HH_{01} & \mathbf 0 &
    \\
    \HH_{10} & \HH_{00} & \HH_{01}
    \\
    \mathbf 0 & \HH_{10} & \ddots
  \end{bmatrix}.
\end{equation}
Calculating the Green function for an arbitrary number of blocks along the semi-infinite
direction follows,
\begin{subequations}
  \begin{align}
    \widetilde{\mathbf Y} &= [\SO_{00} \zz - \HH_{00} - \SE^L]^{-1}(\SO_{01} \zz -
    \HH_{01})
    \\
    \widetilde{\mathbf X} &= [\SO_{00} \zz - \HH_{00} - \SE^R]^{-1}(\SO_{10} \zz -
    \HH_{10})
    \\
    \G_{nn} &= [\SO_{00} \zz - \HH_{00} - \SE^L - \SE^R]^{-1}
    \\
    \G_{mn} & = -\widetilde{\mathbf X}\G_{m-1n} \quad, \text{for $m>n$}
    \\
    \G_{mn} & = -\widetilde{\mathbf Y}\G_{m+1n} \quad, \text{for $m<n$}.
  \end{align}
\end{subequations}
A key-point is that the real-space Green function for a bulk system is a Toeplitz matrix,
e.g. $\G_{mn}=\G_{m'n'}$ for $m - n = m' - n'$. Consequently for a bulk system of $M$
blocks one can calculate the full Green function matrix by only calculating $\G_{n0}$ and
$\G_{nM}$ for all $n$ (omitting $\G_{MM}$ since it equals $\G_{00}$). Thus only $2M - 2$ matrix
multiplications are required in order to calculate the full Green function once $\G_{nn}$,
$\widetilde{\mathbf X}$ and $\widetilde{\mathbf Y}$ are obtained. We note that if the
system is not bulk (e.g. surfaces) this algorithm need only be replaced by the full BTD
algorithm\cite{Papior2017}, which is still much faster than full matrix inversion
algorithms.

\subsection{Bloch's theorem}%
\label{sec:bloch}

We want to obtain the self-energy for the pristine system consisting of a unit cell repeated $N$ times in the transverse direction, large enough to include the defect cf. Fig.~\ref{fig:problem}c. Due to the screening approximation we assume that the potential is unperturbed at the boundary and outside this cell and thus, the self-energy can be calculated from the pristine periodic system. To this end we can apply Bloch's theorem and 
express the $N$ times bigger system transverse to the semi-infinite direction via the primary matrix for a given $\kk$. In our case we are interested in the Green function for a given $\kk$. The equations for expanding the Green function (or any Bloch matrix) along
a single direction for a given $K$ (defined in the large $N$ system) is,
\begin{equation}
  \label{eq:bloch}
  \G_{K}^N =\frac1N
  \;
  \sum_{
      \mathclap{
          \substack{j=0\\
              k_j=\frac{K+j}{N}
          }
      }
  }^{N-1}
  \quad
  \begin{bmatrix}
    1 & \cdots & e^{i (1-N)k_j}
    \\
    e^{i k_j} & \cdots & e^{i (2-N)k_j}
    \\
    \vdots & \ddots & \vdots
    \\
    e^{i (N-1)k_j} & \cdots &1
  \end{bmatrix}
  % \begin{bmatrix}
  %   1
  %   &
  %   e^{-i k_j}
  %   &
  %   \cdots
  %   &
  %   e^{i (1-N)k_j}
  %   \\
  %   e^{i k_j}
  %   &
  %   1
  %   &
  %   \cdots
  %   &
  %   e^{i (2-N)k_j}
  %   \\
  %   \vdots
  %   &
  %   \vdots
  %   &
  %   \ddots
  %   &
  %   \vdots
  %   \\
  %   e^{i (N-1)k_j}
  %   &
  %   e^{i (N-2)k_j}
  %   &
  %   \cdots
  %   &1
  % \end{bmatrix}
  \otimes
  \G_{k_j}^1
\end{equation}
Here $\G_{k_j}^1$ is the primary cell Green function matrix at the primitive cell $k$-point $k_j$ which is to be
unfolded into the matrix $\G_K^N$ and $\otimes$ is the tensor product. The above equation
is only expressed in terms of expansion along one direction, however it is easily
generalized for more than one direction.

% DONE

The above three steps conclude the calculation of the real-space Green function for
arbitrarily large pristine, periodic systems, $N\times M$. The algorithm in short; the self-energies
remove the integral along one $\kk$ direction, the BTD algorithm expands the Green function to arbitrary length, $M$, along the semi-infinite direction employing just matrix multiplications, and
finally Bloch's theorem expands the Green function to arbitrary width (and also  depth for 3D), $N$.

\subsection{Self-energy}%

While the real-space Green function calculates spectral quantities in a pristine system it
is rarely competitive with regular diagonalization methods in the $00$ sub-space and using
Bloch's theorem. Our key mission in calculating the real-space Green function is that it
holds the real-space self-energy, $\rsSE$, which in turn allows truly single defects
(bulk) and contacts (transport) using the Green function
formalism\cite{Brandbyge2002,Papior2017}.

\begin{figure}
  \centering
  \includegraphics{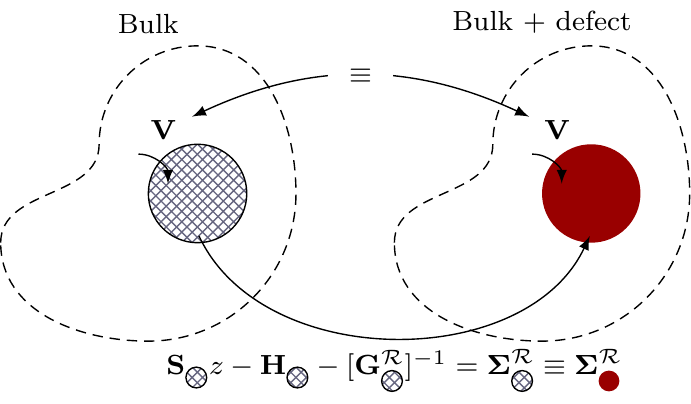}
  \caption{Left system is
      completely bulk and coupled to an internal, also bulk, region (hatched) via $\mathbf V$. Right
      system is a single defect highlighted in color coupled to an infinite bulk region
      via the same $\mathbf V$. Calculation of the real-space self-energy in the colored
      region can be performed by calculating the real-space Green function in the left, bulk,
      system, inverting and subtracting the real-space matrices $\HH$, $\SO$ to retrieve
      the self-energy describing the external bulk part. The resulting $\rsSE$ may be used
      in \emph{any} defected system as long as $\mathbf V$ and the potential in the surrounding region is not changed.
      \label{fig:SE}
  }
\end{figure}

In \Fref{fig:SE} a schematic calculation shows how the real-space Green function may be
used to calculate the real-space self-energy in a region predefined by $M$ and $N$
multiples of the $00$ region as specified in Secs.~\ref{sec:btd} and \ref{sec:bloch}
(hatched region). The real-space self-energy may be conveniently written in two ways:
\begin{align}
  \label{eq:SE-rs-1}
  \rsSE_{00} &= \mathbf V_{00,\mathcal R} \mathbf G^{\mathcal R/00}\mathbf V_{\mathcal
      R,00},
  \\
  \label{eq:SE-rs-2}
  \rsSE_{00} &= \rsS z - \rsH - [\rsG_{00}]^{-1},
\end{align}
where $\mathbf G^{\mathcal R/00}$ is the real-space Green function for the entire bulk
system, excluding the inner region $00$. From \Eref{eq:SE-rs-1} it is clear that
$\rsSE_{00}$ is non-zero only on sites that connects it to the outside through
$\mathbf V_{\mathcal R,00}$. \Eref{eq:SE-rs-2} shows how it is obtained using the real-space Green function.

\section{Results}%

In the following we show results on spectral
and transport properties of truly single defects/junctions
using the real-space self-energy.
Our self-consistent DFT$+$NEGF
is implemented in \siesta, \tsiesta\ and \tbtrans\cite{Soler2002,Papior2017} while the
algorithms described in Sec.~\ref{sec:method} are implemented in \sisl\cite{zerothisisl}.

Three systems will be shown using graphene as the real-space electrode. The different
systems highlight three particular cases where the real-space self-energy is
applicable. We omit the use case of cylindrical self-energies since its use is limited to 3D bulk systems with periodicity along one direction (line defect). The atomic structure of the systems is shown as insets with coloured atoms indicating the support of the real-space self-energy/electrode (in blue), and other electrodes (in
red). A last set of atoms is high-lighted (in light green) which are used as the projection region for local density of states (LDOS) analysis.

All calculations are performed using a $300\,\mathrm{eV}$ mesh cut-off, single-$\zeta$
polarized basis set, and PBE$+$GGA exchange-correlation\cite{Perdew1996}, and otherwise default parameters. Although LCAO
calculations for graphene using simple basis sizes (DZP) misses the lowest unoccupied states\cite{Papior_2018} we do
not add basis orbitals to describe these. Thus our presented analysis is limited to energies below the
missing unoccupied bands ($E-\mu_{\mathrm{graphene}}<3.35\,\mathrm{eV}$).

\subsection{Validation --- Graphene}

\begin{figure}
  \centering
  \includegraphics{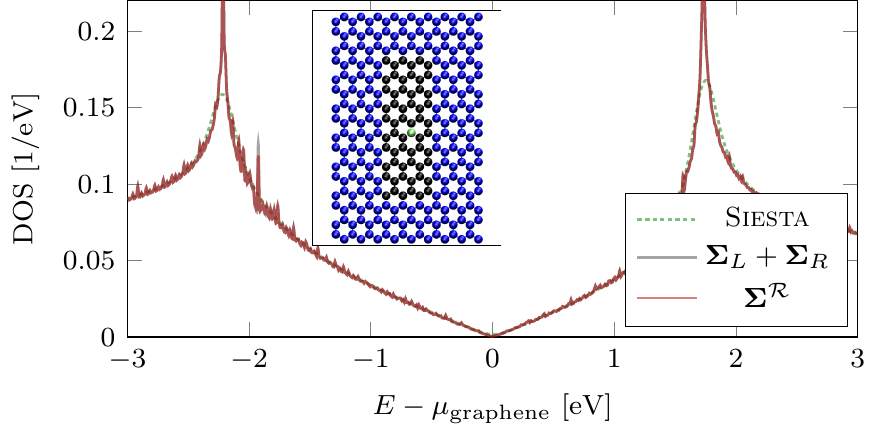}
  \caption{Projected DOS on a single carbon atom in the pristine graphene lattice. Three
      methods are compared (see inset geometry), carbon atom highlighted; \siesta\ PDOS,
      two terminal (bottom/top) \tsiesta\ and finally using $\rsSE$. All methods yield
      exactly the same DOS.
      \label{fig:validity}
  }
\end{figure}

To ensure a functioning method we have constructed a pristine graphene calculation (inset Fig.~\ref{fig:validity}) and calculated the projected DOS on a single carbon atom (marked).
The \siesta\ method calculates the DOS on a $31\times51$ Monkhorst-Pack grid\cite{Monkhorst1976} 
(with energy broadening $\sigma\sqrt2=0.1\,\mathrm{eV}$), and both the Green function methods are based on $300\,\kk$ points
and an imaginary part of $\eta=0.1\,\mathrm{meV}$.

The three different methods all yield the same projected DOS and there is a negligible difference between
the methods.
Any small difference from the diagonalization method vs. the Green function methods lies in the insufficient
$\kk$-point sampling and the large smearing value. The two Green function methods are almost numerically the same
since the system is bulk and no mixing of $\mathbf k$-points take effect.

\subsection{Nitrogen defect}%

% Times
% Siesta: 19187.146 (from good DM) + PDOS

% TS-2: 1885.141 (from good DM)
% TBT-2: 5484.373
%  -2: 7369.513999999999

% GF-TS-1: 143.05
% TS-1: 1364.365 (from good DM)
% GF-TBT-1: 1267.25
% TBT-1: 324.669
%  -1: 3099.334

\begin{figure}
  \centering
  \includegraphics{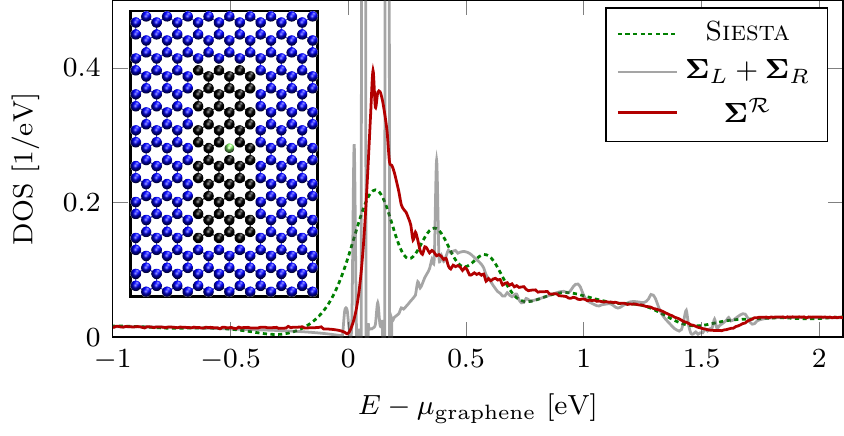}
  \caption{Projected DOS on a single nitrogen defect in the graphene lattice. Three
      methods are compared (see inset geometry), nitrogen atom highlighted; \siesta\ PDOS,
      two terminal (bottom/top) \tsiesta\ and finally using $\rsSE$. The unoccupied states has very
      different character in the three cases.
      \label{fig:nitrogen}
  }
\end{figure}

Single nitrogen defects in graphene intrinsically have a substantial interaction range and
thus calculating defect properties at the DFT level proves difficult \cite{Lambin2012}. In this example we
use the real-space self-energy and compare with a 2D periodic and 1D periodic
calculation. In all 3 examples we use the same unit-cell consisting of a square graphene
lattice cell replicated $8\times9$ totalling 288 atoms.
In \Fref{fig:nitrogen} we show the projected DOS on the nitrogen atom for the three
cases: i) A \siesta\ calculation using a $31\times51$ Monkhorst-Pack grid
in agreement with other work\cite{Hou2013}, ii) a two terminal \tsiesta\ with $300$ transverse
$\kk$ points, and finally iii) using the real-space self-energy calculated from $300$ $\kk$
points. We remark that $300$ $\kk$ points corresponds to $2400$ $\kk$ in the minimal square
graphene unit cell (see e.g. Fig.~\ref{fig:G-2k}).

The DOS shows distinguished differences and particularly so for energies above the graphene Fermi level. The \siesta\ and two probe calculations reveals a fine structure with
multiple peaks dispersed over $\sim1\,\mathrm{eV}$. A large smearing parameter
($\sigma\sqrt2=0.1\,\mathrm{eV}$) for the \siesta\ calculation was required due to the
relatively crude Monkhorst-Pack grid, which still took more than 5 hours on 20 cores. The
two probe calculation shows some even more localized features which could be the same as
those in the \siesta\ calculation. Both look similar to prior calculations\cite{Hou2013}
where the projected DOS on the nitrogen defect ($p_z$) in a similar periodic simulation
was dispersed across two bands with a dispersion $\sim 0.5\,\mathrm{eV}$. We find the real-space method broadens the peaks to a single peak, just above the chemical potential. This result is in perfect agreement with results from a tight-binding description of the isolated N fitted to DFT\cite{Lambin2012}.
Although not shown, the same localized features found for the nitrogen atom are seen for the three neighbouring carbon atoms. These carbon atoms are particularly important for STM images\cite{Lv2012}.

\subsection{STM tip on graphene}%

% Times (TS: 4 k) (TBT: 300 k, 600 E)
% TS-3: 5372.272 (from good DM)
% TBT-3: 31043.080 (300 k, 600 E)
%  -3: 36415.352

% GF-TS-2: 642.2
% TS-2: 5465.622 (from good DM)
% GF-TBT-2: 2966.6
% TBT-2: 248.511
%  -2: 9322.933

Scanning tunnelling microscope\cite{STM} (STM) is a key experimental technique for
analyzing the local electronic structure of surfaces and defects or adsorbates on surfaces. The STM technique is a \emph{single} tip junction probing the spatial
local DOS and yields considerable insight of surface electronic topographies. However, DFT-NEGF
calculations of STM on almost isolated defects are problematic both due to periodic repetition of the surface unit-cell, including the repetition of the STM probe tips. 
Here a calculation of the transmission from an ``STM''-like tip to graphene\cite{Andrei_2012,Bhandari2016,Tetienne2018,VanderHeijden2016,Mark2012} is calculated via two methods. Namely, a three terminal
(left/right graphene/tip) invoking transverse periodicity, and a two terminal (graphene/tip) calculation, both at an applied bias of $\mu_{\mathrm{graphene}}-\mu_{\mathrm{tip}}=-0.5\,\mathrm{eV}$.

\begin{figure}
  \centering
  \includegraphics{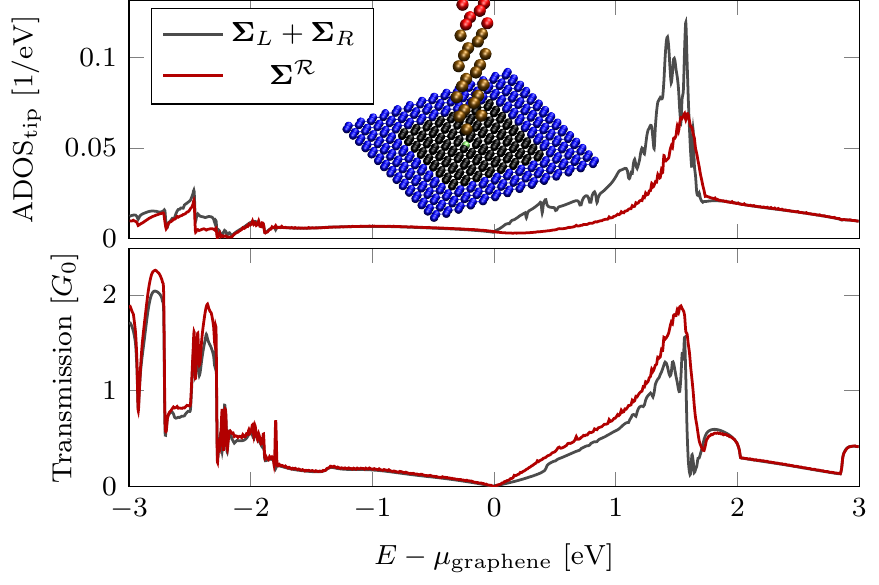}
  \caption{Non-equilibrium transport calculation of tip-graphene contact (tip at $+0.5\,\mathrm V$). Top: spectral DOS from the tip on the first carbon atom in the graphene
      layer. Bottom: Transmission from tip into graphene comparing
      three terminal (left/right/tip) and two terminal (graphene/tip) setups. A large
      difference in the unoccupied energy range is seen both in the spectral DOS and the
      transmission. 
      \label{fig:stm}}
\end{figure}

Figure \ref{fig:stm} top panel shows spectral DOS of scattering states originating from the tip electrode projected on the carbon atom in contact with tip. In the
bottom panel we show the transmission from the tip into graphene.

For the occupied states there is little to no difference while we find a large difference
for the unoccupied states. The spectral DOS decreases on the contact atom while the transmission increases. In both
graphs we find a discontinuity at $0.45\,\mathrm{eV}$ for the 3-electrode simulation (non
existing in the real-space method) which we attribute to periodic image interaction. This
fact is supported by other work\cite{Calogero2019} as well as it matches the bias on the
tip.

Such STM calculations which may be very influential when calculating inelastic
contributions\cite{Halle2018,Garcia-Lekue2011} since they use the energies in the unoccupied
range\cite{Frederiksen2007}.

\subsection{Graphene contacted to a zGNR}

% Times (TS: 3 k) (TBT: 300 k, 600 E)
% TS: 10180
% TBT: 8205.863 (300 k, 600 E)
%  sum: 18385.863

% GF-TS-R: 8923 * 4 / 20 = 1784.6
% TS-R: 2835.845
% GF-TBT-R: 46.87*600 * 4 / 20 = 5624.4
% TBT-R: 156.643
%  sum: 10401.488

A typical experiment comprise large electrodes contacted through a \emph{single} junction
and rarely arrays of contacts present with few exceptions such as e.g. self-assembled
monolayers\cite{dubi2014,schmaltz2017}. A key issue in DFT$+$NEGF simulations of such devices is that, until now, the
simulation had a periodic array of junctions. Such an array of junctions will have
interference effects and requires extra care in convergence of the
width\cite{Thygesen2005} and $\kk$ points. Using the real-space self-energy we eliminate
the periodic junctions and effectively retain a \emph{single} junction where interference
is removed.

\begin{figure}
  \centering
  \includegraphics{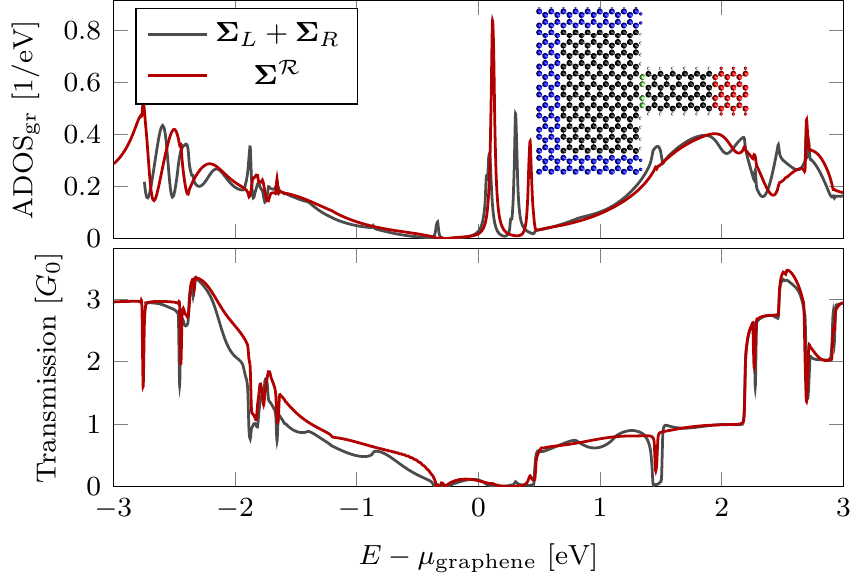}
  \caption{Non-equilibrium transport calculation of graphene-ribbon junction (ribbon at $+0.5\,\mathrm V$). Top: spectral DOS from graphene for the 4 atoms starting in the zGNR ribbon. Bottom:
      transmission between graphene and zGNR.
      A noticeable shift in the localized peaks above the graphene chemical potential and
      some differences in the transmissions. The small differences between the real-space
      method and the standard way is because the system is already relatively wide and
      thus the interference effects are already minor.
      \label{fig:gr2rib}}
\end{figure}

The example shown here is a graphene flake contacted to a zig-zag graphene nano-ribbon
(zGNR) \cite{Leitherer2019}. Our calculations are performed using
$\mu_{\mathrm{graphene}} - \mu_{\mathrm{zGNR}}=-0.5\,\mathrm{eV}$. We remark that any
molecular junction (for instance Au-benzene-di-thiol-Au\cite{Martin2008,Kergueris1999}) could be replaced in this example since the
electrodes are handled as ``surfaces''.
In \Fref{fig:gr2rib} we plot the projected DOS on the first $4$ atoms in the zGNR (top)
and the transmission (bottom). In this example there are relatively few differences since
the unit cell is already relatively wide and thus the interference is limited. There are
however differences such as a larger spread on the localized states just above the
graphene chemical potential. These correspond to states in the zGNR which depends on the electrode coupling and thus is sensitive to periodicities\cite{nick2016,PhysRevB.88.161401}.

\subsection{Electrostatics}

For all results shown, the electrostatics has been solved using the Fourier
transform. Such a method forces the Poisson solution to be periodic. This is in
contrast to our real-space Green function method which inherently has open
boundaries. \tsiesta\ allows external input to ensure the correct boundary conditions for
the electrodes. We have ensured that adding such boundary conditions does not
change the results noticeably, for further information see \cite[Sec.~3.5]{Papior2017}.

\subsection{Disk space}

Currently, our method relies on storing the self-energies on-disk since the algorithm still needs to be implemented in \tsiesta. Each surface self-energy
file has a memory footprint which can be summarized by 3 integer, $n$ (matrix dimension), $n_\kk$ (number of $\kk$-points) and $n_E$ (number of energy points)
\begin{equation}
    \label{eq:disk-usage}
    \operatorname{M}_{\mathrm{GB}}(n,n_\kk,n_E) = 16\cdot n_\kk n^2 (2+n_E) / 1024^3,
\end{equation}
where $2+n_E$ stems from the Hamiltonian and overlap per $\kk$, and the self-energies per $E$. In Tbl.~\ref{tbl:memory} the dimensions and maximum disk usage is shown for the largest bias used.

\begin{table}
  \centering
  \caption{Maximum disk space requirements for the 3 test examples. TS: \tsiesta; TBT: \tbtrans.
      For $\kk$ resolved TBT self-energy calculations one need not store the self-energies for the
      since those can be calculated when needed. For $300$ $\kk$ points they would use $\sim500\,\mathrm{GB}$.
      \label{tbl:memory}
  }
  \begin{tabular}{c|c|c}
    Graphene-Nitrogen & TS & TBT
    \\
    $n$, $n_\kk$ & $n_E = 51$ & $n_E = 600$
    \\
    \hline
    \phantom{$\rsSE$: }$\phantom{2}648$, $6$ & $1.99\,\mathrm{GB}$ & ---
    \\
    $\rsSE$: $2052$, $1$ & $3.33\,\mathrm{GB}$ & $37.8\,\mathrm{GB}$
    \\[2ex]
    Graphene-STM & TS & TBT
    \\
    $n$, $n_\kk$ & $n_E = 178$ & $n_E = 600$
    \\
    \hline
    \phantom{$\rsSE$: }$\phantom{2}720$, $6$ & $8.34\,\mathrm{GB}$ & ---
    \\
    $\rsSE$: $2160$, $1$ & $12.5\,\mathrm{GB}$ & $41.9\,\mathrm{GB}$
    \\[2ex]
    Graphene-zGNR & TS & TBT
    \\
    $n$, $n_\kk$ & $n_E = 178$ & $n_E = 600$
    \\
    \hline
    \phantom{$\rsSE$: }$\phantom{2}648$, $6$ & $6.76\,\mathrm{GB}$ & ---
    \\
    $\rsSE$: $1060$, $1$ & $3.01\,\mathrm{GB}$ & $10.1\,\mathrm{GB}$
  \end{tabular}
\end{table}

We find a required disk space requirement of $\sim50\,\mathrm{GB}$ which is large, but in no way limiting its application on common HPC systems. One generally requires many more energy points in the \tbtrans\ calculation, however since one can define the chemical potential for the real-space electrode to be constant for all applied bias $\mu_{\mathfrak{e}}^{\mathcal R} = 0$ and the other electrode(s) to be at $\mu_{\mathfrak e'} = V$ one can reuse the file for all applied bias' at a much reduced computational cost and with a single file.

\subsection{Performance}%

We have now shown that using the real-space self-energy one can avoid using the super-cell approximation for non-periodic structures such as isolated defects or single junction conductors. In order for it to be competitive with standard methods
it also needs to be competitive in terms of performance/through put. We will here show that
it is in fact less demanding to do a real-space self-energy calculation when taking into
account the full sequence of calculations.

An important factor in using our method is the real-space self-energy calculation. The $\rsSE$
method is slower compared to $\SE_L$/$\SE_R$ given that the self-energy is more costly to
calculate because of larger $\kk$-point sampling and a more complex algorithm. On the
other hand the SCF cycles and transport/DOS calculations are much faster since no
$\kk$-point sampling is required. In Tbl.~\ref{tbl:timing} we show the timings of the presented calculations divided into three
segments; i) \tsiesta, ii) \tbtrans, and iii) $\rsSE$. All timings are based on the same 20-core
machine.
\begin{table}
  \centering
  \caption{Timings of the various steps in the presented calculations, the timings are
      seconds per core in a 20 core setup. TS: \tsiesta; TBT: \tbtrans;
      $\rsSE$: calculating the real-space self-energies for both TS and TBT. All
      calculations are done on the same machine. 
      \label{tbl:timing}
  }
  \begin{tabular}{c|c|c}
    Graphene-Nitrogen & Timing [s] & Total [s]
    \\
    \hline
    \siesta\  $+$ PDOS & $19187$ & $19187$
    \\
    TS $+$ TBT & $1885 + 5484$ & $7369$
    \\
    TS $+$ TBT $+$ $\rsSE$ & $1364 + 324 + 1410$ & $3099$
    \\[2ex]
    Graphene-STM & \phantom{Timing }[s] & \phantom{Total }[s]
    \\
    \hline
    TS $+$ TBT & $5372 + 31043$ & $36415$
    \\
    TS $+$ TBT $+$ $\rsSE$ & $5465 + 248 + 3608$ & $9322$
    \\[2ex]
    Graphene-zGNR & \phantom{Timing }[s] & \phantom{Total }[s]
    \\
    \hline
    TS $+$ TBT & $10180 + 8206$ & $18386$
    \\
    TS $+$ TBT $+$ $\rsSE$ & $2836 + 157 + 7409$ & $10401$
  \end{tabular}
\end{table}

As can be seen, the timings for \tsiesta\ is more or less constant while the \tbtrans\
calculations are much faster. Note however, that for the graphene-zGNR system the
convergence for the real-space method is faster leading to decreased timings in
\tsiesta. Otherwise, the clear bottleneck is the $\rsSE$ calculation which can easily be
embarrassingly parallelized. We remark that, as noted in \Fref{fig:SE}, the self-energy is
generic for \emph{any} defect that does not alter the coupling out to the infinite
exterior. This means that a single calculation of the self-energy allows using it for more
than one system. Since these are one-shot calculations there is no reason not to do an
extremely fine $\kk$ integration when sampling the real-space self-energy. For the systems
shown here it takes less than $100\,\mathrm s$ per energy point for $300$ $\kk$ points. It
should be stressed that the current implementation is done in Python/Cython and thus
additional performance gains would be to port it to fortran/C code.

All-in-all we find that the proposed method is comparable to, or faster, than the existing
method for equivalent $\kk$-point sampling.

\section{Conclusion}%

We have presented a simple, effective, and fast algorithm for constructing real-space self-energies generalized for surfaces and full 2D/3D bulk systems. The algorithm relies on
already well established methods used in the community and can thus be directly integrated
into existing codes without problems. The current algorithms are implemented in the
\tsiesta, \tbtrans\ and \sisl\ toolboxes which are all open-source under GPL variant
licenses.

We have applied the method in three graphene cases which are readily found in current
experimental literature\cite{Lv2012,Andrei_2012,Halle2018,Martin2008,Kergueris1999}. A recurring difference between the analyzed DOS and
transmission profiles is that the occupied energy range is largely comparable to standard
DFT$+$NEGF methods, while the unoccupied energy range shows substantial deviations. Such
differences are attributed to removed interference effects.

We have shown how the use of real-space self-energies will remove the periodic images of
defects in DFT calculations. The results shown provide insights into the far-field
accuracy of DFT$+$NEGF calculations for single defects which has been missing in the
electronic structure community.

\section{Acknowledgments}

We thank Dr.~K.~Kaasbjerg for useful discussions.
Financial support by Villum Fonden (00013340), Danish research council (4184-0003). The
Center for Nanostructured Graphene (CNG) is sponsored by the Danish Research Foundation
(DNRF103).

% \bibliographystyle{apsrev-title}
%\bibliography{refs}

\end{document}